\documentclass[ifip]{svmult}
\usepackage[latin1]{inputenc}
\usepackage[T1]{fontenc}
\usepackage{textcomp}
\usepackage{xspace}
\usepackage[english]{babel}
\usepackage[bottom]{footmisc}
\usepackage{amsmath}
\providecommand{\href}[2]{#2}

\providecommand{\hypersetup}[1]{}
\providecommand{\url}[1]{#1}
\usepackage[dvips]{graphicx}        
\newcommand{\ie}{, {i.e.},\xspace}
\newcommand{\eg}{, {e.g.},\xspace}

\begin{document}

\title*{Trusted Ticket Systems and Applications}
\author{Nicolai Kuntze\inst{1} \and Andreas U. Schmidt\inst{1}}
\institute{Fraunhofer--Institute for Secure Information Technology SIT\\
Rheinstraße 75, 64295 Darmstadt, Germany\\
\texttt{\{andreas.u.schmidt,nicolai.kuntze\}@sit.fraunhofer.de}}
\maketitle
\begin{abstract}
Trusted Computing is a security base technology that will perhaps be
ubiquitous in a few years in personal computers and mobile devices alike.
Despite its neutrality with respect to applications, it has raised
some privacy concerns.
We show that trusted computing can be applied for service access control
in a manner protecting users' privacy.
We construct a ticket system --- a concept which is at the heart
of Identity Management --- relying solely on the capabilities of the trusted platform module
and the standards specified by the Trusted Computing Group.
Two examples show how it can be used for pseudonymous and protected service access.
\end{abstract}
\section{Introduction}\label{intro}
In a ticket based authentication and authorisation protocols like
Kerberos \cite{Kerberos} software tokens are used to prove the identity 
of a single entity. Based on this tokens access to certain systems
is restricted to entities producing appropriate tokens. Additionally
data embodied in the token can be used to implement also an
authorisation control enabling a token based access control scheme beside the mere 
authentication. These tokens are an electronic analog to physical
tickets. They can have a limited validity period ore be used  
a specified number of times.

For the adept, 
some base concepts of Trusted Computing (TC) look very similar to Identity Management (IDM).
We exploit the analogy between TC and IDM and construct a ticket system using TC functionality,
thus obtaining a cornerstone of IDM systems.
The applicability of such a \textit{trusted ticket system}
is shown in the context of a reputation system and a push service.

Section~\ref{sec:TCess} provides necessary background on TC.
Section~\ref{sec:TC-Tickets}, developing the trusted ticket system proper is
subdivided in~\ref{sec:AIKTickets}, explaining how to use Attestation Identity Keys (AIKs) 
for the realisation of tickets, and~\ref{sec:AcqRed}
detailing the proceedings for the acquisition and redemption of the latter.
Section~\ref{sec:GenArc} outlines a general service access architecture
utilising trusted tickets, with a high degree of separation of duties, 
providing for pseudonymity,
accountability, and charging functionality.
Section~\ref{sec:App} embeds trusted ticket systems in the two mentioned application
contexts and discusses resulting benefits.
Conclusions are drawn in Section~\ref{sec:conclusions}.
\section{Trusted Computing Essentials}\label{sec:TCess}
Trusted computing uses a hardware anchor as a root of trust and is now entering the mobile domain 
with the aim to provide a standardised  security infrastructure. 
Trust in the context of
TC means (as defined by the Trusted Computing Group, TCG) that an entity  always behaves in the expected manner for the 
intended purpose. 
The trust anchor, called Trusted Platform Module (TPM), 
offers various functions related to security.

Each TPM is bound to a certain environment and together they form a trusted platform (TP)
from which the TPM cannot be removed. 
Through the TPM the TP gains a cryptographic engine and a
protected storage. 
Each physical instantiation of a TPM has a unique identity by an 
Endorsement Key (EK) which is created at manufacture time. 
This key is used as a base for secure 
transactions as the Endorsement Key Credential (EKC) asserts that the holder of the private portion
of the EK is a TPM conforming to the TCG specification. 
The EKC is issued as well at production 
time and the private part of the key pair does not leave the TPM. 
There are other credentials specified by the TCG 
which are stating the conformance of the TPM and the platform for instance the so called 
platform credential.
Before a TPM can be used a take ownership procedure must be performed in which the usage of 
the TPM is bound to a certain user.
The following technical details are taken from~\cite{TPM06}. 

The TPM is equipped with a physical random number generator, and a key generation
component which creates RSA key pairs. The key generator is designed as a
protected capability and the created private keys are kept in a shielded
capability (a protected storage space \textit{inside} the TPM). 

The TPM possesses so-called shielded capabilities protecting internal
data structures by controlling their use. Three of them are essential for applications.
First, \textit{key creation and management}, 
second the ability to create a \textit{trust measurement} 
which can be used to assert a certain state toward a, remote party, and finally
methods, which we call \textit{sealing}, to protect arbitrary data by binding
it (in TCG nomenclature) to TP states and TPM keys.  

For the TPM to issue an assertion about the system state,
a process called \textit{attestation}, two protocols are available. 
As the uniqueness of every TPM leads to privacy concerns, they provide
pseudonymity, respectively, anonymity. 
Both existing attestation protocols rest on Attestation Identity Keys (AIKs) which are placeholders for the EK. 
An AIK is a 1024 bit RSA key the private portion of which is also sealed inside the TPM. 
The simpler protocol of Remote Attestation (RA) offers pseudonymity by introducing a 
trusted third party, the privacy CA (PCA, see~\cite{TCGIWG1}), which issues a credential
stating that the respective AIK is generated by a sound TPM within a
valid platform. 
The system state is measured by a reporting process with the 
TPM as its central reporting authority receiving the measurement values and calculating a
unique representation of the state using hash values. 
For this the TPM has several Platform Configuration registers (PCR).
Beginning with the system boot each component reports a measurement value, e.g., a hash value over
the BIOS, to the TPM and stores it in a log file. During RA the communication
partner which acts as verifier receives this log file and the corresponding PCR value.
The verifier can then decide if the device is in a configuration which is 
trustworthy from his perspective.
Apart from RA, the TCG has defined Direct Anonymous Attestation. This involved protocol is
based on a zero knowledge proof but due to certain constraints of the hardware it is not
implemented in current TPMs.

\begin{figure}[tbp]
  \centering 
  \includegraphics[width=0.4\textwidth]{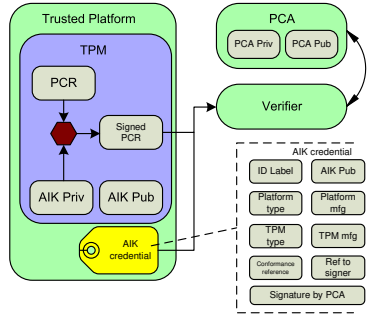}
  \caption{Remote Attestation process.}
  \label{fig:Remote_Attestation}
\end{figure}

AIKs are crucial for applications since they can not only be used, 
according to TCG standards, to attest the  origin and authenticity of a trust measurement, 
but also to authenticate other keys generated by the TPM. 
Before an AIK can testify the authenticity of any data, a PCA 
has to issue a credential for it.
This credential together with the AIK can therefore be used as an identity
for this platform.  
The protocol for issuing this credential consists in three basic steps. 
First, the TPM generates an RSA key pair by performing the \verb+TPM_MakeIdentity+
command. The resulting public key together with certain credentials  identifying
the platform is then transferred to the PCA. Second, the PCA 
verifies the correctness of the produced credentials and the AIK signature. If
they are valid the PCA creates the AIK credential which contains an identity
label, the AIK public key, and information about the TPM and the platform.
A special structure containing the AIK credential is created which is
used in step three to activate the AIK by executing the \verb+TPM_ActivateIdentity+
command. So far, the TCG-specified protocol is not completely secure, since
between steps two and three, some kind of handshake between PCA and platform is
missing. The existing protocol could  sensibly 
be enhanced by a challenge/response part to
verify the link between the credentials offered in step one and used
in step two, and the issuing TPM. The remote attestation process is shown in 
figure \ref{fig:Remote_Attestation}.

Beside the attestation methods TC offers a concept to bind
blobs of data to a single instantiation and state of a TPM. 
The \verb+TPM_unbind+ operation
takes the data blob that is the result of a \verb+Tspi_Data_Bind+ command and
decrypts it for export to the user. 
The caller must authorise the use of the key that will
decrypt the incoming blob. In consequence this 
data blob is only accessible if the platform is in the namely 
state which is associated with the respective PCR value.

A mobile version of the TPM is currently being defined by the TCG's
Mobile Phone Working Group~\cite{MTM09}. This so-called Mobile Trusted Module 
(MTM) differs significantly from the TPM of the PC world and is in fact more powerful in some respects.
In particular, it contains a built-in verifier for attestation requests,
substituting partly for an external PCA.
Both TPM and MTM are a solid basis for application architectures and scenarios.
Trusted Computing not only affects the world of networked PCs but will
also heavily impact the mobile industry. Accordingly, one of our application scenarios
see section \ref{sec:Push} is a mobile one. The concept for ticket system we present in section
\ref{sec:TC-Tickets} is in fact network agnostic and can be applied to the Internet as well 
the mobile world. Further mobile scenarios can be found in \cite{KuntzeSchmidt2006A,KuntzeSchmidt2006C}.
\section{A TC-based Ticket System}\label{sec:TC-Tickets}
The basic idea is to establish a (pseudonymous) ticket system using the
identities embodied in the PCA-certified AIKs. Specific about our design is
the tickets are generated locally on the (mobile) device of the user.
Ticket acquisition an redemption rests solely on trusted computing methods
implemented in the TPM chip embedded in the users platform.
We first describe how AIKs can be turned into tickets that can be used
in a ticket-based service access or identity management architecture,
and then develop the processes for their acquisition and redemption.
\subsection{AIKs as Tickets}\label{sec:AIKTickets}
For security considerations the TPM restricts the usage of AIKs. 
It is not possible to use AIKs as signing keys for arbitrary data and in particular 
to establish tickets in that way.
It is therefore necessary to employ an indirection using a TPM generated signing key
and certify this key by signing it with an AIK --- \textit{viz} \textit{certify} it
in the parlance of the TCG. 
Creation of a key is done by executing the
\verb+TPM_CMK_CreateKey+ command, which returns an asymmetric key pair where the
private portion is encrypted by the TPM for use within the TPM only. 
The resulting
key pair is loaded into the TPM by \verb+TPM_LoadKey+
and thereafter certified by \verb+TPM_CertifyKey+. 
By certifying a specific key the TPM makes the statement that
``this key is held in a TPM-shielded location, and it will never be
revealed''. For this statement to have veracity, a challenger or verifier  must
trust the policies
used by the entity that issued the identity and the maintenance policy of
the TPM manufacturer. 

This indirection creates to each AIK a certified key (by the namely AIK)
that can be used for signing data, in particular the payload data 
of a ticket to be submitted to, and accepted by, a service.
We call this key pair the \textit{certified signing key} (CSK).
CSK, AIK, together with a certificate by the PCA (see below)
attesting the validity of that AIK, are the ingredients that realise a ticket
for a single operation\eg a service access.
\subsection{Ticket Acquisition and Redemption}\label{sec:AcqRed}
Tickets are acquired by a \textit{trusted agent} (TA)\ie the user of a
ticket system and associated services operating with his trusted platform, from the PCA.
They are then redeemed at the \textit{(ticket) receiving system} (RS). 
In both processes, a \textit{charging provider} (CP) may occur as a third party, 
depending on application architectures. 
We now describe how these operations proceed.

Note that we do not distinguish between public and private key portions of a
certificate establishing a credential. As a notation, 
the credential of some certified entity 
$\text{Cert(\textit{entity}, \textit{certificate})}$ means the union of
the public key $\text{Pub(\textit{certificate})}$ and the entity signed
with the certificate's private key, $\textit{entity}_{\text{Priv}(\textit{certificate})}$.
Verifying a credential means to check this digital signature.

An interesting option is that
the credentials issued by the PCA for a AIKs can be designed as \textit{group credentials}\ie
they do not identify a single AIK \textit{viz} ticket but rather its price or
value group $g$ chosen from a predetermined set indexed by the natural numbers $g\in\{1,\ldots,G\}$.
The group  replaces an individual identity of a platform and many TAs will get the
same group certificate. Only the PCA can potentially resolve the individual identity
of a platform.
This allows for the combination of a value proposition with privacy protection, as
these groups are used to implement price and value discrimination of tickets.
Note that the PCA is free in the choice of methods to implement group certificates.
This could be done by simply using the same key pair for the group or by the existing,
sophisticated group signature schemes~\cite{Chaum91}.

\begin{figure}
\centering 
  \resizebox{0.95\textwidth}{!}{\includegraphics{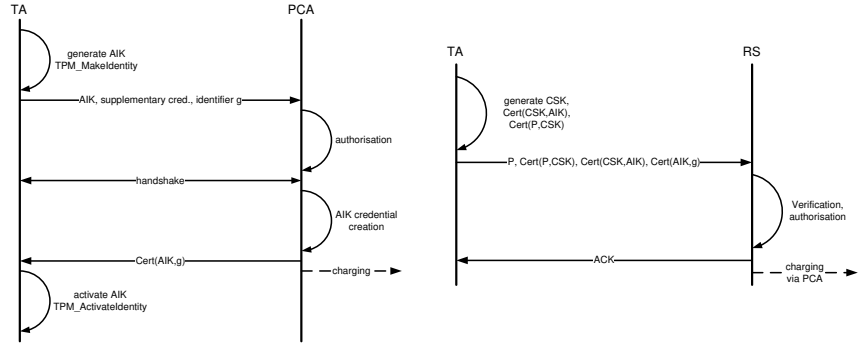}}
  \caption{Ticket acquisition (left) and redemption (right) processes.}
  \label{fig:Ticket_Acquisition_Redemption}
\end{figure}

If a TA wants to acquire a rating ticket from group $g$, he first generates an 
AIK using the \verb+TPM_MakeIdentity+ command.
Next, TA requests from the PCA a credential for this AIK, belonging to group $g$,
by sending AIK, group identifier and supplementary data as required by TCG protocols,
to the PCA.
The PCA now knows the identity of the TA.
This can be used to perform a charging for the ticket, 
either by contacting CP or by the PCA itself
(how charging actually works is not in the scope of this paper).
It is important that, at this stage, an authorisation decision on the ticket generation
can be made by the PCA, for instance to blacklist misbehaving participants.
If the authorisation succeeds (and not earlier, to save bandwidth and resources), 
the PCA performs a handshake operation with the TA to ensure that the AIK has actually
been generated by the particular TPM in question.
Upon success, the PCA generates the credential
$\text{Cert}(\text{AIK},g)$ certifying that the AIK belongs to group $g$.
The credential is transferred back to TA, where finally the 
\verb+TPM_ActivateIdentity+ command is executed to enable subsequent usage of this AIK.
The process is shown on the left hand side of Figure~\ref{fig:Ticket_Acquisition_Redemption}.

Redeeming a ticket is now very simple, as shown on the right
hand side of Figure~\ref{fig:Ticket_Acquisition_Redemption}.
TA has first to generate a CSK\ie a public/private key pair and
the credential $\text{Cert}(\text{CSK},\text{AIK})$ for it according
to the process described in Section~\ref{sec:AIKTickets}.
He then signs a certain payload, $P$\eg describing a service request, 
with CSK to obtain $\text{Cert}(\text{P},\text{CSK})$.
The payload and the credential chain
$\text{Cert}(\text{P},\text{CSK})$, $\text{Cert}(\text{CSK},\text{AIK})$, 
$\text{Cert}(\text{AIK},g)$
is then transferred from TA to RS and this set of data embodies the ticket 
proper (we do not discuss a particular data format for the ticket).
RS verifies this chain and makes an authorisation decision, for instance
to implement a protection against multiple spending.
Finally, RS acknowledges receipt of $P$ and optionally
initiates another charging operation (ex post charging) via PCA.
\subsection{A Generic Architecture for Service Access}\label{sec:GenArc}
\begin{figure}
\centering 
  \resizebox{0.6\textwidth}{!}{\includegraphics{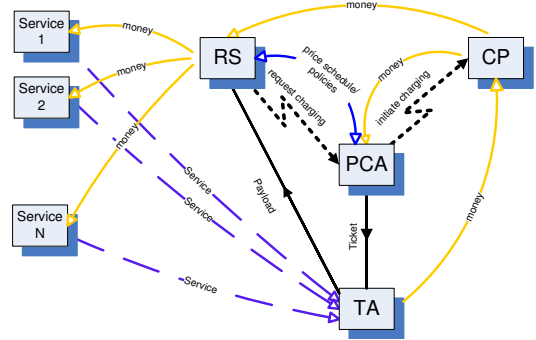}}
  \caption{A generic architecture using TC-based tickets.}
  \label{fig:Gen_Arch}
\end{figure}

The embedding of the described ticket acquisition and redemption into an application 
system and business context offers many variants.
A very basic scenario is shown in Figure~\ref{fig:Gen_Arch}.
Here, a trusted agent (user) would like to access some service, 
and buys a ticket of a certain value from the PCA.
The ticket belongs to a certain group which can represent statements such as ``for usage with 
Service $n$'', and a certain value, monetary or intangible\eg in a rebate scheme.
The user then issues a service request as payload in the ticket 
redemption toward RS.
The TA pays for the ticket at the CP at the time of redemption of the ticket and the
CP distributes revenue shares between himself,
PCA, and RS, according to service level agreements. RS, in turn, remunerates the service
(an acknowledgement of service processing is omitted for simplicity).

This realises an access control scheme to multiple services mediated by PCA and RS,
yielding three essential benefits: 1.~non-repudiation by the chain of credentials, 
2.~accountability by resolution of the TA's identity through PCA, 
and 3.~pseudonymity by separation of duties.
The PCA/RS combination plays a very central role for the control of identities embodied
in the pseudonymous tickets that PCA issues. It is in fact an embodiment of the
role of an identity provider in a ticket-based identity management (IDM) system.
That TC can be used to model IDM was outlined in~\cite{KuntzeSchmidt2006A,KuntzeSchmidt2006C}, 
and is presented here for the first time in detail.

Though the separation of duties between PCA and CP allows in principle even
for anonymity of the person using a TA, since only upon charging
this person must be identified by credit card account or other means, this
may not be the best option.
In fact this would loose some accountability of the TA users.
While RS may be able to obtain personal identities from PCA
if pertinent contractual relationships are in place\eg if fraud by a TA user is suspected,  
data protection  regulations may prevent a CP from unveiling personal
identities.
A second role played by the PCA is the initiation of charging.
With respect to the revenues from ticket sales, a natural approach
is a sharing between RS and PCA (and CP for its service).
RS and PCA negotiate and implement policies for authorisation within
the ticket acquisition and redemption processes\eg to prevent double spending
or to blacklist misbehaving users.
In collaboration between PCA and RS, practically any price schedule can be realised. 

This architecture naturally extends to an arbitrary number of receiving systems
to which PCA offers ticket management, and in particular pricing, as services.
A further extension would be to let TA express values of 
tickets by using different (groups of) CSKs. In this way tickets can
be associated with additional certified attributes\eg priorities.
\subsection{Security and Privacy}\label{sec:Sec}
The presented method for the management of tickets
provides for perfect pseudonymity of the participants toward the system.
In fact, only PCA is able to de-anonymise users.

Note that for the namely reason only PCA can initiate a charging, since only
he knows (or is able to know) the identity of a TA and can link
it to the identity of the corresponding participant.
To keep this pseudonymity strength, it is essential that our concept relies
only on genuine TPM functionality, and in particular avoids the usage
of trusted software.
If there was a trusted software managing tickets in some way at the side of TA, 
then this software, and the state of the platform would have to be attested both
in ticket acquisition and redemption.
To this end the TC protocols for remote attestation transfers trust measurements and
measurement logs to the corresponding verifier (PCA or RS in our case).
These data can however --- and this is a principal problem with remote attestation ---
be used to individualise the trusted platform, if, as in the PC domain,
the number of system states and different measurement logs created at boot time,
is very large in relation to the number of users of a TC-based service.
Besides, avoiding remote attestation saves bandwidth and resources consumption. 
Since in this case no trust can be laid in the TA for ticket management,
some kind of double, or multiple, spending protection or 
usage authorisation is needed at RS upon ticket redemption.

On the other hand security necessitates additional means of protection of content in transit.
In many applications,
for instance if confidentiality of transported payload is a protection target,
 trusted software usage cannot be avoided
We show in Section~\ref{sec:Push} how TC  can be used to establish end-to-end protection
for $P$, but this definitely requires trusted software clients at both ends.
\section{Two Applications}\label{sec:App}
\subsection{Price Scheduling in Pseudonymous Rating Systems}\label{sec:Rating}
This application has been outlined in~\cite{RepuTicketsVG}.
Electronic market places for physical and 
information goods are increasingly occupied by self-organising
communities. These market places exhibit the characteristics of the
so-called long tail economy~\cite{longTail}. That is, the classical asymmetry between 
suppliers and consumers is lifted. Buyers and sellers are often even in numbers
and may change their roles dynamically. Virtual, or physical, 
goods are offered in large numbers and diversity and with potentially small
demand for each single one.
Matchmaking and orientation of buyers is difficult in a long tail economy,
long term relationships are hard to build, and trust between trade partners
must be established somehow~\cite{Bakos1998}.

A common approach is to let market players themselves provide the necessary 
guidance. This is mostly embodied in  reputation systems by which buyers and 
sellers rate each other and the goods sold, or recommendation systems\ie  
programs which attempt to predict items that a user may be interested in, 
given some information about the user's profile.
Reputation systems, according to Paul Resnick \textit{et al.}~\cite{Resnick2000}
``seek to establish the shadow of the future [the expectation of reciprocity
or retaliation in future interactions, cf.~\cite{Axelrod1984}] 
to each transaction by creating an expectation that other people will look back on it''. 
The goal is to establish a homogeneous market for honest participants.
That community ratings (of goods) do in fact strongly influence buyer 
behaviour is shown empirically in~\cite{SDW06}. 

Existing reputation systems are fragile, in that they can easily 
be distorted or abused even within the frame of laws governing them. 
`Attacks' of this kind, though not proper attacks in the sense of 
information security, threaten the integrity --- with respect to its purpose ---
of the informational content stored in the system.
Dellarocas~\cite{Dellarocas00} classifies unfair behaviour 
into the categories
1.~\textit{Ballot stuffing}: A seller colludes with a group of buyers in order to 
be given unfairly high ratings.
2.~\textit{Bad-mouthing}: Sellers and buyers collude to rate other sellers
unfairly low to drive them out of the market.
3.~\textit{Negative discrimination}: Sellers provide good services only to
a small, restricted group of buyers.
4.~\textit{Positive discrimination}: Sellers provide exceptionally good service to
some buyers to improve their ratings.
A situation of controlled anonymity in which the market place knows the identity of
participants and keeps track of all transactions and ratings, but conceals the identity
of buyers and sellers, is identified as essential to avoid unfair behaviour.
For instance, anonymity is an effective protection against bad-mouthing, but 
cannot work for ballot stuffing as sellers can give hidden indications of their
identities to colluders.

On the other hand, the best known individual attack on reputation systems
uses Sybils to obtain a disproportionately large influence~\cite{Douceur2002}.
Friedman and Resnick~\cite{Friedman2001} point to the general problem of 
`cheapness' of pseudonyms in marketplaces and reputation systems, since with name 
changes dishonest players easily shed negative reputation, as corroborated 
theoretically in~\cite{Dellarocas2004}. The paper~\cite{Bhatta2005} gives an explicit threshold 
for the transaction costs for reputations needed to avoid ballot stuffing. 
However, an indiscriminate pricing
of identities for the submission of ratings poses an undesired entry deterrent.
It seems therefore plausible that reputation systems should be based on pseudonyms
which allow for a flexible forward pricing.

While related work addresses particular vulnerabilities~\cite{Friedman05}, or proposes general
frameworks to ensure accountability in reputation systems while maintaining
anonymity~\cite{Buttyan1999,Zieglera2006}, we here propose a simple mechanism to
introduce arbitrary costs for pseudonyms.
The separation of duties between PCA and RS in our 
ticket system implements here
precisely the controlled anonymity desired for reputation systems
through the properties 1.--3.\ mentioned in Section~\ref{sec:GenArc}.

Again, the generic ticket system can be embedded in various ways into a 
(commercial) reputation system.
If a TA user wants to express a rating about another user 
(for example, a buyer about a seller, a seller about a buyer), he buys 
a  \textit{rating ticket} from the PCA.
The group attribute of the ticket in this special context expresses a value proposition
for the rating\eg an impact factor used by the rating system to calculate weighted overall ratings,
as well as an attribution to a particular rating system.
The user then formulates the rating and sends it to RS as ticket payload, and charging is executed.
The result would be a rating statement about another participant of the 
rating system which is trustworthy, accountable, 
but protected as a pseudonym.
This enables the resolution of one important problem in reputation systems, 
namely accountability of users\ie the possibility to trace back malicious 
ones and threaten them with consequences.
Based on the trusted ticket system, price schedules can be adapted to the requirements
of rating systems as laid out above.
On the extreme ends of the spectrum are cost-free registration of 
ratings by PCA, ensuring only accountability, and increasing charges with the
number of ratings (or\eg their frequency).
Even reverse charging\ie paying incentives for ratings\eg such of good quality,
is possible.
\subsection{Content Protection for Push Services}\label{sec:Push}
Workers occupied with `nomadic' tasks 
depend on infrastructures for easy and swift access to required data. 
E-Mail push services like RIM's Blackberry conquer this market with huge success, 
and aim at high availability and ease of use. 
Push services are characterised by the ability to notify end users of new content. 
For an e-mail service the end user device is activated by the mail 
server, gets new mail, and notifies the user.
The basic method for this has been formulated for instance in the
standards of the Open Mobile Alliance (OMA, see~\cite{OMAPush2006}).
Some providers have extended their range to enable access to
company databases and implement loosely coupled work-flows incorporating nomadic workers.

Due to the high value of the exchanged data these systems are threatened by, 
even professional, attackers, raising the requirement to protect the
distribution of pushed data.
Security concerns can be  grouped in two main areas. 
First, data has to be protected in transit to the device, and 
confidentiality is to be maintained. 
Second, after the data is delivered it has to 
be protected against unauthorised access. 
The latter problem is of practical importance in 
use cases like the mentioned e-mail push, but also for 
SMS delivery, and other data synchronisation processes 
between a central data base and a mobile device.
Current approaches to this challenge are predominantly using software 
tokens\eg PKCS\#7, to secure message transport and storage. 
Such solutions however suffer from the drawback that an attacker can extract keys from memory 
during encryption or decryption of a data block. 
Smart cards are a more evolved approach,
but in the mobile domain no standard has become prevalent.

\begin{figure}
\centering
\includegraphics[width=0.4\textwidth]{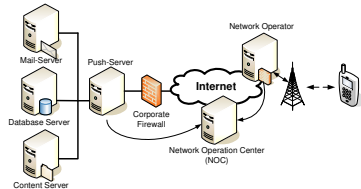}
\caption{A centralised architecture}
\label{NOC}
\end{figure}
Figure 
\ref{NOC} illustrates the widely used, centralised push architecture. In its centre 
a Network Operation Center (NOC) performs all tasks
regarding the communication to the mobile devices. 
The data destined for the mobile devices are stored in sources
like mail servers. These sources excite the push server and deliver the
data. Due to this 
activation the push server either requests a communication line to the
mobile device managed by the NOC or delivers the data to the NOC which 
in turn stores the data until they are handed to the mobile device.
From a company's view the management costs are low as there 
is no additional effort needed to maintain\eg a special firewall configuration.
The decentralised counterpart includes direct communication between push server and mobile device,
requiring access over\eg the Internet to servers behind a company firewall, in turn necessitating
special protection of the internal infrastructure.
Taking a malicious service provider into consideration
privacy concerns are added to the general ones
regarding transport security.
Potentially the NOC can access
every message which is sent to a mobile device, allowing for information leakage. 
First and foremost the message content
could be extracted and disclosed. 
Moreover, analysis of the collaboration between active users becomes possible. 
Both attacks have to be 
treated in the protocol design to enable end-to-end security and to
conceal all sensitive information. 

\begin{figure}
\centering
\includegraphics[width=\textwidth]{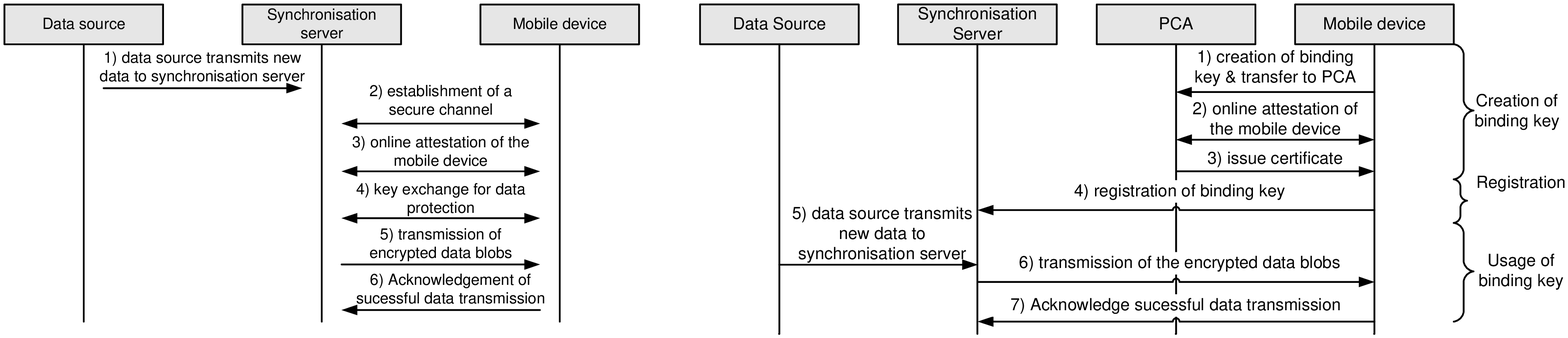}
\caption{Scenarios based on  blob sealing (left) resp.\ key sealing (right)}
\label{graphs}
\end{figure}
Protection of content is a major use case for trusted computing~\cite{MPWGUC05}.
We first present basic options to protect push content, and then describe the 
integration with th trusted ticket system of Section~\ref{sec:TC-Tickets}.
Based on blob binding (see Section~\ref{sec:TCess}) the simplest scenario 
see the left diagram in Figure~\ref{graphs}. 
First, the data source signals the synchronisation server (SyncS) which then
locates the target device, establishes communication, 
and controls the synchronisation relying on\eg
OMA data synchronisation. 
Channel security can be realised using Transport Layer Security. 
Remote attestation of the device and key exchange for payload encryption are performed.
The data is transferred to the device and
stored in sealed  blobs, see~\cite[Chapter~12]{TPM06}, 
only accessible in the determined state of this unique device.
This approach suffers from latency produced 
by the Steps~2-4, in particular
attestation creates computational load and produces some traffic. 

An approach to meet this  challenge is 
shown on the right hand side of Figure~\ref{graphs}. 
SyncS encrypts the data with a public part of a key pair. The use
of the corresponding private part is restricted by a PCR value. 
To grant trust in this public key the SyncS requires a 
certificate issued by a PCA, and a certified PCR value, signifying\eg 
the presence of an e-mail application in 
a certain configuration and a well defined environment.
The PCA which certifies the platform keys can actually be 
used here to issue certificates augmented by the information of the PCR value. 
This variant of an AIK is further called a \textit{binding key}. 
Step~1 transmits the public portion of a key pair to the  designated
PCA for content encryption keys. 
Remote attestation is performed and a certificate is created 
stating that the key originates from a TP and is
usable if and only if the platform is in a certain state. 
This certificate is transmitted to the mobile device, and then, together with the
binding key, to SyncS. 
Steps~1 to~4 only have to be performed once during roll out of the device or take 
ownership by its user. 
In Step~5 the data is transmitted to SyncS, which 
can now encrypt the data with the binding key or with a hybrid scheme. 
Data transmission is executed in Steps~5--7.
Note that the first presented method is much more flexible than the second one, since in the former
the server can decide each time
if the particular device can be considered as trustworthy.

The combination of the ticket system with the content protection system
proceeds as follows. After it has been activated, 
the mobile device uses a ticket obtained from the PCA
as a token to access SyncS which can either be the RS itself or an associated service. 
The ticket AIK cannot be used for data encryption, but the ticket payload can 
carry the mentioned public key for content encryption, \textit{and} its certificate.
The ticket PCA can act concurrently as the CA certifying the binding key or in separation from it.
Ticket grouping or prioritising can be used advantageously in such a scheme\eg
for attribution of bandwidth to a request and load balancing.
\section{Conclusions}\label{sec:conclusions}
We have shown how to generate and use tickets based on TC,
and have provided theoretical proof-of-concept 
in two independent application scenarios
employing trusted tickets.
What we have constructed is essentially a payment system with a trusted 
third party guaranteeing pseudonymity.
It is therefore worthwhile to compare our method with the use case scenario ``Mobile payment''
of TCG's Mobile Phone Working Group Use Case Scenarios~\cite[Section 8]{MPWGUC05}.
There, the focus lies on device-side support of payment operations on a mobile
phone which is turned into a trusted platform. This always involves a trusted
software on the device which is not required in our approach.
On the other hand this is only possible through the introduction of a trusted 
third party, the PCA with its extended duties.
Thus we lack the universality of client-side solutions.
Yet we have shown that a very simple ticket system with strong pseudonymity
can be established resting solely on the most basic TPM functions.

It should be noted that our applications 
use TC in a way very different from Digital Rights Management (DRM), 
which is often considered as the sole use for TC. 
Both applications bind the economic value to a particular 
instantiation of the TPM. 
If this trust anchor breaks, 
only a limited damage can occur as the damage is restricted in space and time\eg
to a single data synchronisation in a push service or submission of a single reputation. 
In contrast, if a single TPM in a DRM system breaks, 
the protected digital good can be converted into an unprotected version 
which can be freely distributed on a large scale, causing heavy monetary losses to its owner.

It is interesting to note that introducing TC yields a secondary user
bound identification token. Due to the take ownership procedure of the TPM
it is bound to a certain user. Therefore it is in its function very 
similar to a SIM as it is also possible to migrate the relevant parts 
from one TPM to the next. One may ask whether two different identification tokens 
will survive in future TC-enabled mobile devices.
%


%
\providecommand{\noopsort}[1]{} \providecommand{\singleletter}[1]{#1}


\begin{thebibliography}{10}
\bibitem{Kerberos}
Massachusetts Institute of Technology:
Kerberos: The Network Authentication Protocol.
http://web.mit.edu/kerberos/

\bibitem{TPM06}
Trusted Computing Group:
\newblock {TCG TPM} specification version 1.2 revision 94.
\newblock Technical report, TCG (2006)

\bibitem{TCGIWG1}
Trusted Computing Group:
\newblock {TCG Infrastructure Working Group Reference Architecture for
  Interoperability (Part I) Specification Version 1.0 Revision 1}.
\newblock Technical report, TCG (2005)

\bibitem{MTM09}
Trusted Computing Group:
\newblock {TCG Mobile Trusted Module Specification. Specification version 0.9
  Revision 1}.
\newblock Technical report, TCG (2006)

\bibitem{Chaum91}
Chaum, D., van Heyst, E.:
\newblock Group signatures.
\newblock In Davies, D., ed.: Advances in Cryptology - EUROCRYPT '91. Volume
  547 of Lecture Notes in Computer Science, Berlin, Heidelberg,
  Springer-Verlag (1991)  257--265

\bibitem{KuntzeSchmidt2006A}
Kuntze, N., Schmidt, A.U.:
\newblock Transitive trust in mobile scenarios.
\newblock In Müller, G., ed.: Proceedings of the International Conference on
  Emerging Trends in Information and Communication Security (ETRICS 2006).
  Volume 3995 of Lecture Notes in Computer Science (LNCS), Springer-Verlag
  (2006)  73--85

\bibitem{KuntzeSchmidt2006C}
Kuntze, N., Schmidt, A.U.:
\newblock Trusted computing in mobile action.
\newblock In Venter, H.S., Eloff, J.H.P., Labuschagne, L., Eloff, M.M., eds.:
  Proceedings of the Information Security South Africa (ISSA) Conference (2006)

\bibitem{RepuTicketsVG}
Kuntze, N., Schmidt, A.U.:
Employing Trusted Computing for the forward pricing of pseudonyms in reputation systems.
To appear in: Proceedings of the Workshop Virtual Goods at the Conference AXMEDIS
2006, Leeds, England, 13.--15.~December~2006

\bibitem{longTail}
Anderson, C.:
\newblock The Long Tail. Wired Issue 12.10 - October 2004
http://web.archive.org/web/20041127085645/http://www.wired.com/wired/ archive/12.10/tail.html

\bibitem{Bakos1998}
Bakos, Y.:
\newblock The emerging role of electronic marketplaces on the internet.
\newblock Commun. ACM \textbf{41} (1998)  35--42

\bibitem{Resnick2000}
Resnick, P., Kuwabara, K., Zeckhauser, R., Friedman, E.:
\newblock Reputation systems.
\newblock Communications of the ACM \textbf{43} (2000)  45 -- 48

\bibitem{Axelrod1984}
Axelrod, R.:
\newblock The Evolution of Cooperation.
\newblock Basic Books, New York (1984)

\bibitem{SDW06}
Salganik, M.J., Dodds, P.S., Watts, D.J.:
\newblock Experimental study of inequality and unpredictability in an
  artificial cultural market.
\newblock Science \textbf{311} (2006)  854--856

\bibitem{Dellarocas00}
Dellarocas, C.:
\newblock Immunizing online reputation reporting systems against unfair ratings
  and discriminatory behavior.
\newblock In: ACM Conference on Electronic Commerce. (2000)  150--157

\bibitem{Douceur2002}
Douceur, J.R.:
\newblock The sybil attack.
\newblock In Druschel, P., Kaashoek, F., Rowstron, A., eds.: Peer-to-Peer
  Systems: First InternationalWorkshop, IPTPS 2002 Cambridge, MA, USA, March
  7-8, 2002. Volume 2429 of Lecture Notes in Computer Science, Springer-Verlag
  (2002)  251--260

\bibitem{Friedman2001}
Friedman, E.J., Resnick, P.:
\newblock The social cost of cheap pseudonyms.
\newblock Journal of Economics \& Management Strategy \textbf{10} (2001)
  173--199

\bibitem{Dellarocas2004}
Dellarocas, C.:
\newblock Sanctioning reputation mechanisms in online trading environments with
  moral hazard.
\newblock MIT Sloan Working Paper No. 4297-03 (2004)

\bibitem{Bhatta2005}
Bhattacharjee, R., Goel, A.:
\newblock Avoiding ballot stuffing in ebay-like reputation systems.
\newblock In: P2PECON '05: Proceeding of the 2005 ACM SIGCOMM workshop on
  Economics of peer-to-peer systems, New York, NY, USA, ACM Press (2005)
  133--137

\bibitem{Friedman05}
Cheng, A., Friedman, E.:
\newblock Sybilproof reputation mechanisms.
\newblock In: P2PECON '05: Proceeding of the 2005 ACM SIGCOMM workshop on
  Economics of peer-to-peer systems, ACM Press (2005)  128--132

\bibitem{Buttyan1999}
Buttyan, L., Hubaux, J.P.:
\newblock Accountable anonymous access to services in mobile communication
  systems.
\newblock In: Proceedings of the 18th IEEE Symposium on Reliable Distributed
  Systems. (1999)  384--389

\bibitem{Zieglera2006}
Zieglera, G., Farkas, C., L{õ}rincz, A.:
\newblock A framework for anonymous but accountable self-organizing
  communities.
\newblock Information and Software Technology \textbf{48} (2006)  726--744

\bibitem{MPWGUC05}
Trusted Computing Group:
\newblock {M}obile {P}hone {W}orking {G}roup {U}se {C}ase {S}cenarios - v 2.7.
\newblock Technical report, TCG (2005)

\bibitem{OMAPush2006}
{Open Mobile Alliance}:
\newblock Push architecture. draft version 2.2 - 20 jan 2006.
  oma-ad-push-v2\_2-20060120-d.
\newblock Technical report, Open Mobile Alliance (2006)

\end{thebibliography}
\end{document}